\documentclass[prb,aps,epsf,twocolumn,superscriptaddress,showpacs,10pt]{revtex4-2}
\usepackage{graphicx,amsfonts,times,bm,amsmath,verbatim,color,array}
\usepackage{lineno}
\usepackage{xcolor}
\usepackage{algorithmic}
\usepackage[colorlinks,urlcolor=blue,citecolor=blue,linkcolor=blue]{hyperref}

\begin{document}

\title{Quantum Hall Effect in a Weyl-Hubbard Model: Interplay between Topology and Correlation}

\author{Snehasish Nandy}
\affiliation{Theoretical Division, Los Alamos National Laboratory, Los Alamos, New Mexico 87545, USA}
\affiliation{Center for Nonlinear Studies, Los Alamos National Laboratory, Los Alamos, NM, 87545, USA}

\author{Christopher Lane}
\affiliation{Theoretical Division, Los Alamos National Laboratory, Los Alamos, New Mexico 87545, USA}

\author{Jian-Xin Zhu}
\affiliation{Theoretical Division, Los Alamos National Laboratory, Los Alamos, New Mexico 87545, USA}
\affiliation{Center for Integrated Nanotechnologies, Los Alamos National Laboratory, Los Alamos, New Mexico 87545, USA}

\begin{abstract}
 The interplay between topology and electronic correlations offer a rich avenue for discovering emergent
quantum phenomena in condensed matter systems. In this work, starting from the Weyl-Hubbard model, we
investigate the quantum Hall effect to explore the consequence of onsite Hubbard repulsion on nontrivial Weyl
band topology in the presence of an external magnetic field. Within the Gutzwiller projected wavefunction
method, we find the system to undergo multiple topological phase transitions by tuning on-site Coulomb
interaction, including two distinct Weyl phases with different numbers of Weyl node pairs and a trivial narrow
band insulator. Crucially, these two Weyl phases may be identified by the sign of their chiral Landau levels.
The possible experimental signature of these topological phases and correlation effects is provided by the
magnetic-field dependent quantum Hall conductivity within the Kubo response theory. 
\end{abstract}

\maketitle

\section{Introduction} The three-dimensional (3D) Weyl semimetal (WSM) has been of great interest in the condensed-matter community over the last decade due to its unique non-trivial band topology. WSMs, which emerge from breaking either spatial inversion (IS) or time-reversal (TR) symmetries or both simultaneously, are characterized by hosting Weyl nodes in the bulk generated by momentum space touching of nondegenerate valence and conduction bands at isolated points~\cite{Murakami_2007, Peskin_1995, Nagaosa_2007, Ran_2011, Balents_2011, Leon_2011, Zhong_2011, Sergey_2011}. The topological properties of WSMs are manifested in the fact that these Weyl nodes act as the source and sink of Abelian Berry curvature, and are protected by a nontrivial integral Chern number $C=\pm 1$, which is related to the strength of the magnetic monopole enclosed by the Fermi surface~\cite{Mele_2018, Niu_2010}. As a consequence, the WSMs host topologically protected surface states, so called Fermi arcs, that connect Weyl nodes of opposite monopole charges. According to the ``no-go" theorem, the Weyl nodes in WSM come in pairs of positive and negative monopole charges (also called chirality) and the net monopole charge summed over all the Weyl nodes in the Brillouin zone exactly vanishes~\cite{Ninomiya_1981, Ninomiya_1983}.

After the discovery of the WSM phase in real materials, e.g., $\rm TaAs$ family and $\rm WTe_2$, significant attention has been devoted to understanding the theoretical and experimental properties induced by the non-trivial topology at the single particle level~\cite{Moore_2018,Yan_2017, Mele_2018, Ding_2021, Liao_2017, Ong_2021}. Moving beyond the single particle paradigm by including electron-electron correlation effects brings about an astonishingly rich and complex set of phases including unconventional superconductivity~\cite{Keimer_2015} and colossal magnetoresistance~\cite{Dagotto_2003}, so the question arises ``How does non-trivial band topology compliment or compete with correlation effects in quantum matter?" In this connection, several recent works have explored the interplay between Weyl-band topology and electronic correlations. Specifically, it has been proposed that intermediate electronic correlations can give rise to flat bands in WSMs~\cite{Xu_2020}, whereas strong electron-electron interactions can gap out the bulk Weyl nodes, thus precipitating a phase transition towards either a Weyl-Mott insulator~\cite{Morimoto_2016}, an axion insulator~\cite{ Roy_2015, Ali_2012,Kim_2012}, a topological superconducting phase~\cite{Haldane_2018}, a pair-density wave phase related to space-time supersymmetry~\cite{Jian_2015}, or a Weyl-CDW phase~\cite{Shou_2013, Shi_2021, Sekine_2014, Wang_2016,Roy_2016,Laubach_2016}. Another possible consequence of electronic correlations is the emergence of a Weyl-Kondo semimetal, which has recently been experimentally realized in $\rm YbPtBi$~\cite{Guo_2018}, $\rm RAlGe$ compounds (with R$= \rm La$ and $\rm Ce$)~\cite{Corasaniti_2021} and  Ce$_3$Bi$_4$Pd$_3$~\cite{Paschen_2017, Dzsaber_2013}. Very recently, a new route has been proposed to
significantly enhance the dark matter detection efficiency via
strongly correlated topological Weyl semimetal in the absence
of external magnetic field~\cite{Huang_2023}. However, despite these vigorous efforts in just the last few years, very little has been done to examine the signature of the correlated WSM phase in the presence of an external magnetic field.

The topological WSMs exhibit a plethora of intriguing transport phenomena due to their unique band topology in the presence of external fields, which makes the magneto-transport is one of the most powerful methods to probe its band topology~\cite{Yan_2017, Mele_2018, Ding_2021, Liao_2017, Ong_2021,Nandy_2017,Nandy_2020,Spivak_2016, Nandy_2021, Nandy_2022, Zyuzin_2017, Tewari_2016, Tewari_2022}. The magneto-transport in the strong-field limit in WSMs has attracted intensive attention of late due to its underlying Landau level (LL) characteristics. In particular, a 3D quantum Hall effect (QHE) induced by the LLs is predicted to occur in WSMs since the Fermi arcs at the top and the bottom surfaces form a closed loop via ``wormhole" tunneling assisted by the Weyl nodes, and therefore, serving as a direct experimental probe of Weyl band topology~\cite{Xie_2017,Xie_2020,Nag_2022}. Remarkably, the 3D QHE has been realized recently in a non-interacting Dirac semimetal Cd$_3$As$_2$~\cite{Timo_2018, Narayan_2019,Uchida_2017}. In light of the above discussions, it is natural to ask what will happen to the Landau level physics and related transport phenomena in a correlated WSM. 
 
In this article, we investigate the quantum Hall effect in an IS and TR broken Weyl-Hubbard (WH) system to explore the effect of onsite Hubbard Coulomb repulsion on the nontrivial Weyl band topology in the presence of an external magnetic field. By employing the Gutzwiller approximation to treat the electronic correlations, we find the WH system to exhibit multiple topological phases, including two Weyl phases with different pairs of Weyl nodes and a trivial narrow band insulator by tuning on-site Coulomb interaction. Interestingly, in the presence of an external magnetic field, we show the chiral Landau levels to change sign while crossing between Weyl phases. We calculate the magnetic-field dependent quantum Hall conductivity (QHC) within the Kubo response theory to explore possible signatures of the topological phase transitions and correlation effects. Our results on QHC can be directly validated by experiments. The recent discovery of correlated magnetic WSMs, such as $\rm Co_3Sn_2S_2$~\cite{Morali_2019, Chen_2019, Hasan_2019} and $\rm Pr_2Ir_2O_7$~\cite{Won_2021}, provide a platform to experimentally verify our predictions.

\section{Model Hamiltonian of Weyl-Hubbard System} The Weyl-Hubbard semimetal on a cubic lattice (lattice constant $a=1$) can be written as~\cite{Huang_2023}
\begin{eqnarray}
H&&=\sum_{j,ss^\prime}[-t\sigma_{x,ss^\prime}(c^{\dagger}_{js}c_{j+\hat{x},s^\prime}+c^{\dagger}_{js}c_{j+\hat{y},s^\prime}+c^{\dagger}_{js}c_{j+\hat{z},s^\prime}) \nonumber \\
&&-it^\prime(\sigma_{y,ss^\prime}c^{\dagger}_{js}c_{j+\hat{y},s^\prime}+\sigma_{z,ss^\prime}c^{\dagger}_{js}c_{j+\hat{z},s^\prime})+\text{H.c.}] \nonumber \\
&&+m\sum_{j,ss^\prime} \sigma_{x,ss^\prime}c^{\dagger}_{js}c_{js^\prime}+U\sum_{j}n_{j\uparrow}n_{j\downarrow},
\label{WH_Ham}
\end{eqnarray}
where $t, t^{\prime}$ are the hopping parameters, $s, \, s^\prime$ are the spin indices, and $U$ is the onsite Hubbard-interaction strength between two electrons carrying opposite spins. Here, $m$ denotes site energy which acts as an effective in-plane Zeeman term and the first two terms of the Eq.~(\ref{WH_Ham}) represent the kinetic energy part of the Hamiltonian $\rm{H}_{\rm kin}$. In the non-interacting limit (i.e., $\rm{U}=0$), the model Hamiltonian represents a both time-reversal symmetry and inversion symmetry breaking Weyl semimetallic phase containing linearly dispersing Weyl nodes. Specifically, $m$, the effective in-plane magnetic field breaks TR symmetry and the imaginary hopping parameter $t^\prime$ breaks inversion symmetry. Though inversion is formally broken, all the Weyl nodes of the system lie at the same energy.  
\begin{figure}[htb]
\centering
\begin{center}
\includegraphics[width=0.49\textwidth]{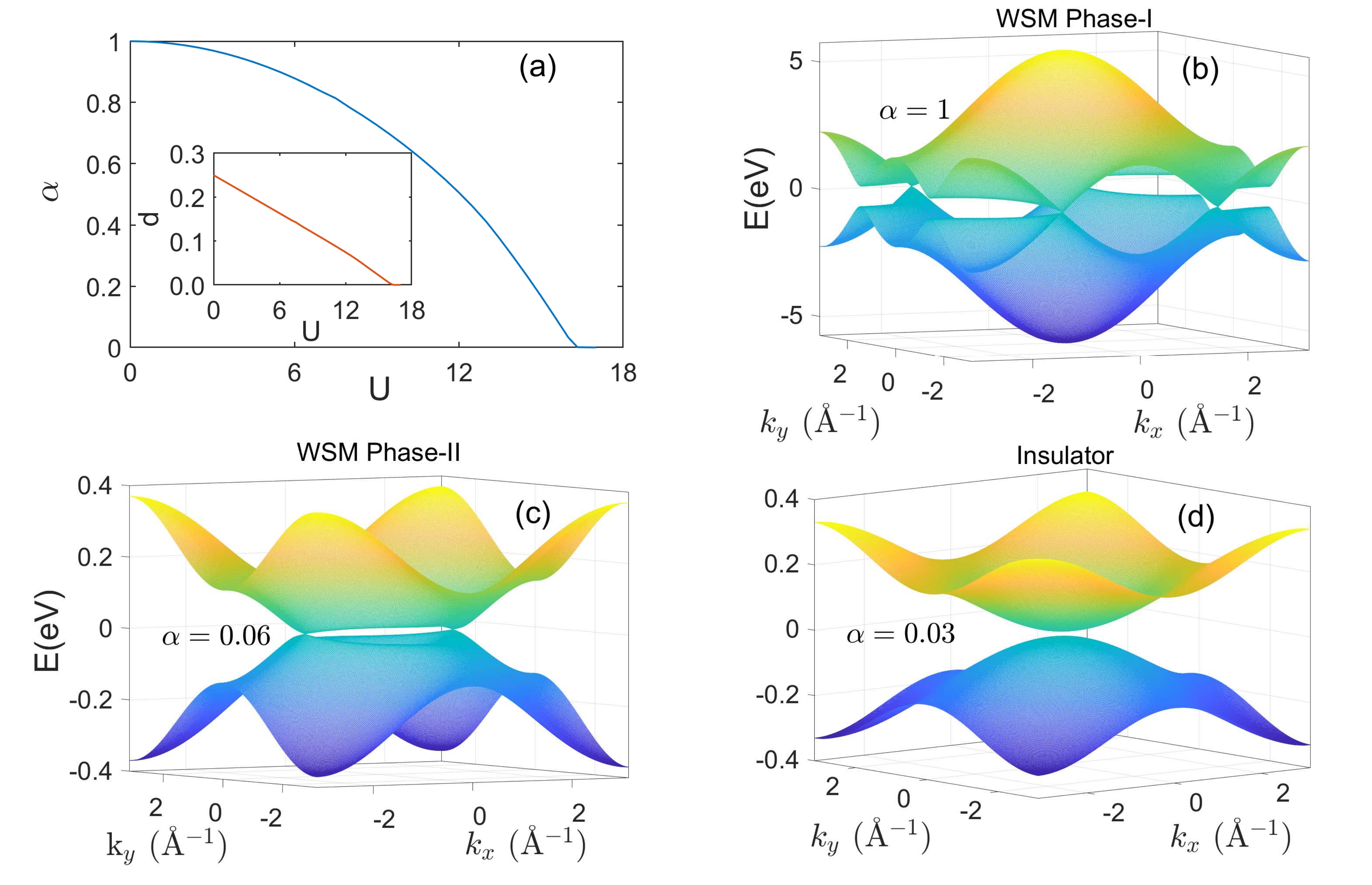}
\caption{(a) The phase diagram $\alpha$ vs $\rm U$ (inset: $d$ vs $\rm U$ plot) of WH sytem is shown. (b)-(d) depict the 3D energy dispersion of different phases (WSM phase-I, WSM phase-II and narrow-band insulator) (k$_z$ is suppressed) during the phase transition. In this figure, the Weyl cones are at $[\pm \cos^{-1}(m/2t_{\rm nor}), \pi, 0]$ and $[\pm \cos^{-1}(m/2t_{\rm nor} -2), 0, 0]$ for WSM phase-I and WSM phase-II respectively. Here, we choose to measure energy in units of $t$:  $t=1$, $t^\prime=0.3\,t$ and $m=0.25\, t$. Here, $\alpha=1,\, 0.06, \, 0.03$ corresponds to $U \sim 0, \, 15.8 \, t, \, 16.2 \, t$ respectively.}\label{fig1}
\end{center}
\end{figure}
We now investigate the effect of onsite Coulomb interactions using the Gutzwiller method approximation, which subjects to the following self-consistency equations at half-filling 
\begin{eqnarray}
\alpha=16(\frac{1}{2}-d)d \quad   {\rm and} \quad U+ \frac{\partial \alpha}{\partial d} \langle H_{\rm kin} \rangle =0,
\label{scf_eqn}
\end{eqnarray}
where $d$ and $\alpha$ are the double occupancy and renormalization factor, respectively. The phase diagram obtained by solving the above self-consistency equations is shown in Fig.~\ref{fig1}(a). Interestingly, we find that, with increasing the $U$ value, the WH system undergoes two topological phase transitions: (i) from WSM phase-I to WSM phase-II at $U \sim 15.4 \, t \, \rm eV$ and (ii) from WSM phase-II to trivial narrow band insulator at $U \sim 16 \, t \, \rm eV$. These two Weyl phases are characterized by the ratio of $m$ and renormalized hopping parameter $t_{\rm nor}=\alpha \, t$, in particular, $m/2t_{\rm nor} < 1$ and $m/2t_{\rm nor} > 1$ for Weyl phase-I and Weyl phase-II respectively~\cite{Huang_2023}. The topology of the different phases is uncovered via the Berry curvature analysis. Specifically,
calculating positive and negative monopole charge as well as
their location distinguishes two different Weyl phases. On the
other hand, we reveal the insulating phase is $Z_2$ topologically
trivial via Wilson loop analysis.

In particular, after turning on the onsite correlations, the hopping parameters $t, t^{\prime}$ are renormalized by a factor $\alpha$ and consequently, the location of the Weyl nodes changes via the expression $[\pm \cos^{-1}(m/t_{\rm nor}), \pi, 0]$ and $[\pm \cos^{-1}(m/2 t_{\rm nor}), 0, \pi]$. In particular, the nodal points of opposite chirality approach each other with increasing $U$ and finally annihilate at $U \sim 15.4 \, t \, \rm eV$. Then the system enters into the strongly TR broken WSM phase (WSM phase-II) with two Weyl nodes located at $[\pm \cos^{-1}(m/2t_{\rm nor} -2), 0, 0]$ by satisfying the condition ($m/2t_{\rm nor}>1$). If we further increase $U$, the Weyl nodes become gapped and the system encounters the second phase transition from WSM phase-II to a narrow band insulating phase. The energy dispersion of these three phases are shown in Fig.~\ref{fig1}(b)-(d). It is important to note that the velocity at the nodal points are controlled by the renormalized imaginary hopping parameter $t^{\prime}_{\rm nor}$ ($=\alpha t^\prime$). We note that the increase of $U$ values corresponds to a decrease of the kinetic energy, which could be driven by a negative pressure experimentally.

\section{Landau Level Spectrum in the Correlated Regime} To investigate the effect of strong electron correlations on the quantum Hall transport properties of the Weyl-Hubbard semimetal, we implement the external magnetic field (B) contribution via the standard Peierls substitution: $t_{ij}c_i^{\dagger}c_j \rightarrow t_{ij}e^{-i\frac{e}{\hbar}\int_{i}^{j}\mathbf{A(r) \cdot dr}}c_i^{\dagger}c_j$ where $\mathbf{A}$ is the vector potential. We consider the Gutzwiller projected wavefunction approach to treat
the strong correlation effect by reducing the statistical weight of double occupation. Within the Gutzwiller projected wavefunction method~\cite{Wang_2006, Noboru_2008, Zhu_2012}, the renormalized Weyl-Hubbard
Hamiltonian can be written as
\begin{eqnarray}
&& H= \nonumber \\
&&\sum_{j,ss^\prime}\{-t \left(e^{-iB_zj_y}\sqrt{\alpha}_{\mathbf{j-x}s}c^{\dagger}_{\mathbf{j-x} s}+e^{iB_zj_y}\sqrt{\alpha}_{\mathbf{j+x}s}c^{\dagger}_{\mathbf{j+x} s}\right) \nonumber \\
&&+m c^{\dagger}_{js}\sigma_{x,ss^\prime} -it^\prime\left(\sqrt{\alpha}_{\mathbf{j-y}s}c^{\dagger}_{\mathbf{j-y} s}-\sqrt{\alpha}_{\mathbf{j+y}s}c^{\dagger}_{\mathbf{j+y} s}\right)\sigma_{y,ss^\prime}
\nonumber \\
&&-it^\prime \left(e^{iB_xj_y}\sqrt{\alpha}_{\mathbf{j-z},s}c^{\dagger}_{\mathbf{j-z} s}-e^{iB_xj_y}\sqrt{\alpha}_{\mathbf{j+z},s}c^{\dagger}_{\mathbf{j+z} s}\right)\sigma_{z,ss^\prime} \nonumber \\
&&-t[
\left(e^{iB_xj_y}\sqrt{\alpha}_{\mathbf{j-z}s}c^{\dagger}_{\mathbf{j-z} s}+e^{iB_xj_y}\sqrt{\alpha}_{\mathbf{j+z}s}c^{\dagger}_{\mathbf{j+z} s}\right)+ \nonumber \\
&&(\sqrt{\alpha}_{\mathbf{j-y}s}c^{\dagger}_{\mathbf{j-y} s}+\sqrt{\alpha}_{\mathbf{j+y}s}c^{\dagger}_{\mathbf{j+y} s})]\sigma_{x,ss^\prime}\} c_{\mathbf{j}s^\prime}+ Ud N_L.
\label{Peierls_sub}
\end{eqnarray}
The site-dependent renormalization parameter $\alpha_{js}$ is given by
\begin{equation} 
\sqrt{\alpha_{js}}=\biggl{[} \frac{(\bar{n}_{js}-d_{j})(1-\bar{n}_j-d_j)}{\bar{n}_{js} (1-\bar{n}_{js})} \biggr{]}^{1/2}+ \biggl{[}\frac{d_{j}(\bar{n}_{j\bar{s}}-d_j)}{\bar{n}_{js} (1-\bar{n}_{js})} \biggr{]}^{1/2} \;,
\end{equation}
with $\bar{n}_{js}$ the expectation value of the number operator $n_{js}=c^{\dagger}_{js}c_{js}$ and $\bar{n}_j=\sum_{s} \bar{n}_{js}$. It is important to note that we find,
in the noninteracting limit, the occupation (or carrier density) and the double occupancy of each site to be the same in the presence of magnetic field, which motivates us to consider the renormalization factors to be homogeneous (site independent), denoted by $\alpha$. Therefore, $\bar{n}_{js} = \bar{n}_{s} = \bar{n}_{\bar{s}}, d_{j} = d$, and $\alpha_{js} = \alpha$. In the above Hamiltonian, the B-field lies within the $xz$-plane, such that it can be represented by the vector potential $\bm{ A}=(-yB_z, 0, yB_x)$ in Landau gauge, yielding $\bm{B}=\bm{\nabla\times A} = B_z\hat{\bm{z}}+B_x\hat{\bm{x}}$. In the following, energy and length are measured in units of $t$, and the cubic lattice constant $a$, respectively. Both $t$ and $a$ are assumed to be one unless specified otherwise. It is clear in the Hamiltonian that both $k_x$ and $k_z$ are good quantum numbers. To satisfy the $y$-direction periodicity, the magnetic field strength is restricted to $2\pi/Q$ where Q is commensurate with $L_y$ such that $Q = L_y/m$ reduces to an integer only. Here, $L_y$ denotes the number of sites along $y$-direction. We also note that the effect of external magnetic field on the renormalization parameter $\alpha$, is negligible. Therefore, we use the values of $\alpha$ and $d$ obtained from the zero-field calculation throughout the rest of this work. 

\begin{figure}[htb]
\centering
\begin{center}
\includegraphics[width=0.49\textwidth]{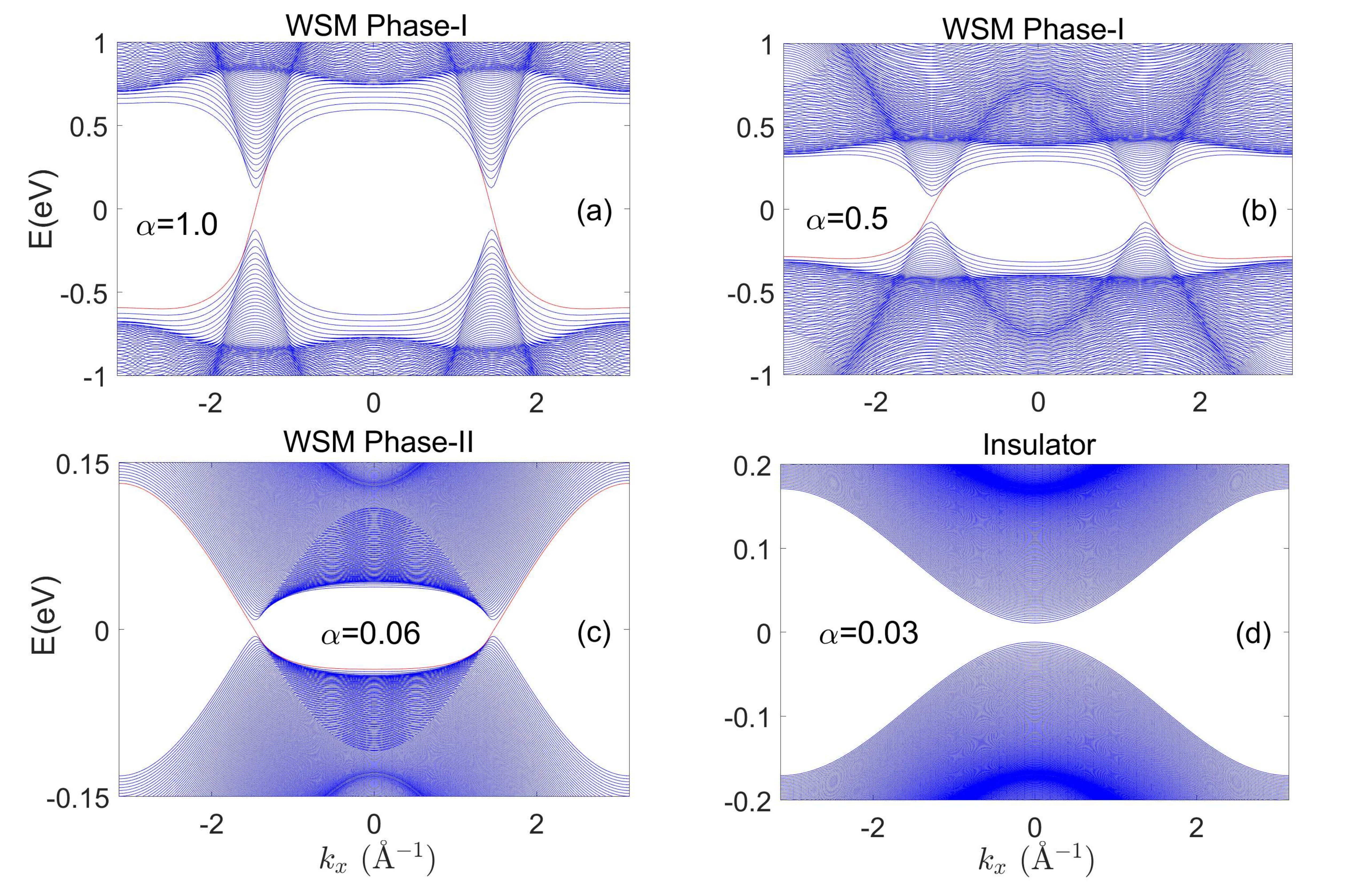}
\caption{(a)-(d) depict the evolution of the LL dispersion of different phases (WSM phase-I, WSM phase-II and narrow-band insulator) as a function of $k_x$ during the phase transition. Here, the external magnetic field is applied along the $x$-direction. The red line in LL spectrum indicates the chiral Landau levels. We have chosen $k_z=0$, $L_y=200$, $B_x=B_z=B_0=2\pi/L_y$. The other parameters we choose to measure energy in units of $t$:  $t=1$, $t^\prime=0.3\,t$ and $m=0.25\, t$. Here, $\alpha=1, \, 0.5, \, 0.06, \, 0.03$ corresponds to $U \sim 0, \, 12 \, t, \, 15.8 \, t, \, 16.2 \, t$ respectively.}\label{fig2}
\end{center}
\end{figure}
  
The evolution of the Landau Level spectrum of WH system for various values of $U~(\alpha)$ obtained by diagonalizing the above Hamiltonian is shown in Fig.~\ref{fig2}. Here, we apply the external $\rm B$ parallel to the separation of the Weyl nodes of opposite chiralities  (i.e., $\mathbf{B} \parallel \hat{x}$). In the WSM phase-I ($\alpha=1$ and $0.5$), a pair of doubly degenerate chiral modes ($n=0^{th}$ LL with $n$ is the Landau level index) clearly appear in the system traversing across the Weyl nodes at $k_x=\pm \cos^{-1}(m/2 \, t_{\rm nor})$, with positive and negative slopes with respect to the applied field direction. The slope of the chiral LLs is determined by the monopole charge of the Weyl node, where a positive (negative) monopole charge gives rise to a chiral mode with a positive (negative) slope. The degeneracy of the Landau levels arises from the conservation of monopole charge. As onsite correlations $\rm U$ are introduced, the LLs flatten while maintaining characteristic band features. Figure~\ref{fig2}(c) shows the LL spectrum of WSM phase-II with two Weyl nodes. Similar to WSM phase-I, a pair of chiral LLs with opposite slopes traverse across the Weyl nodes at $k_x=\pm \cos^{-1}(m/2 \,t_{\rm nor} -2)$. In contrast to WSM phase-I, however, the LLs are non-degenerate in this case. Interestingly, the chiral LLs change sign (slope) during the phase transition from WSM phase-I to WSM phase-II. Finally in the large-$U$ limit [Fig.~\ref{fig2}(d)], the LL spectrum is gapped, indicative of a correlated insulating phase, and all the LLs are non-degenerate. Clearly, chiral LLs do not exist in this phase, due to the gapping out of the Weyl nodes. It is important to note that the LL spectrum in each case is independent of the value of $k_z$.

Furthermore, we would like to point out that when $\rm B$ is applied parallel to the $z$-axis, the counter propagating chiral LLs in each WSM phase cross each other linearly at $k_z = 0$ within the bulk gap of achiral LLs, since they lie on the same momentum projection axis. Compared to the case $\mathbf{B} \parallel \hat{x}$, the main difference is that the bulk LLs are doubly degenerate for both WSM phases irrespective of momentum $k_x$. It is important to note that if we increase the strength of $\rm B$ by integer multiple, $n$, of $B_0=2\pi/L_y$ for a fixed $\rm U$, the degeneracy of the LLs will increase $n$-fold due to the Brillouin zone folding along the $y$-direction.

\begin{figure}[b]
\centering
\begin{center}
\includegraphics[width=0.49\textwidth]{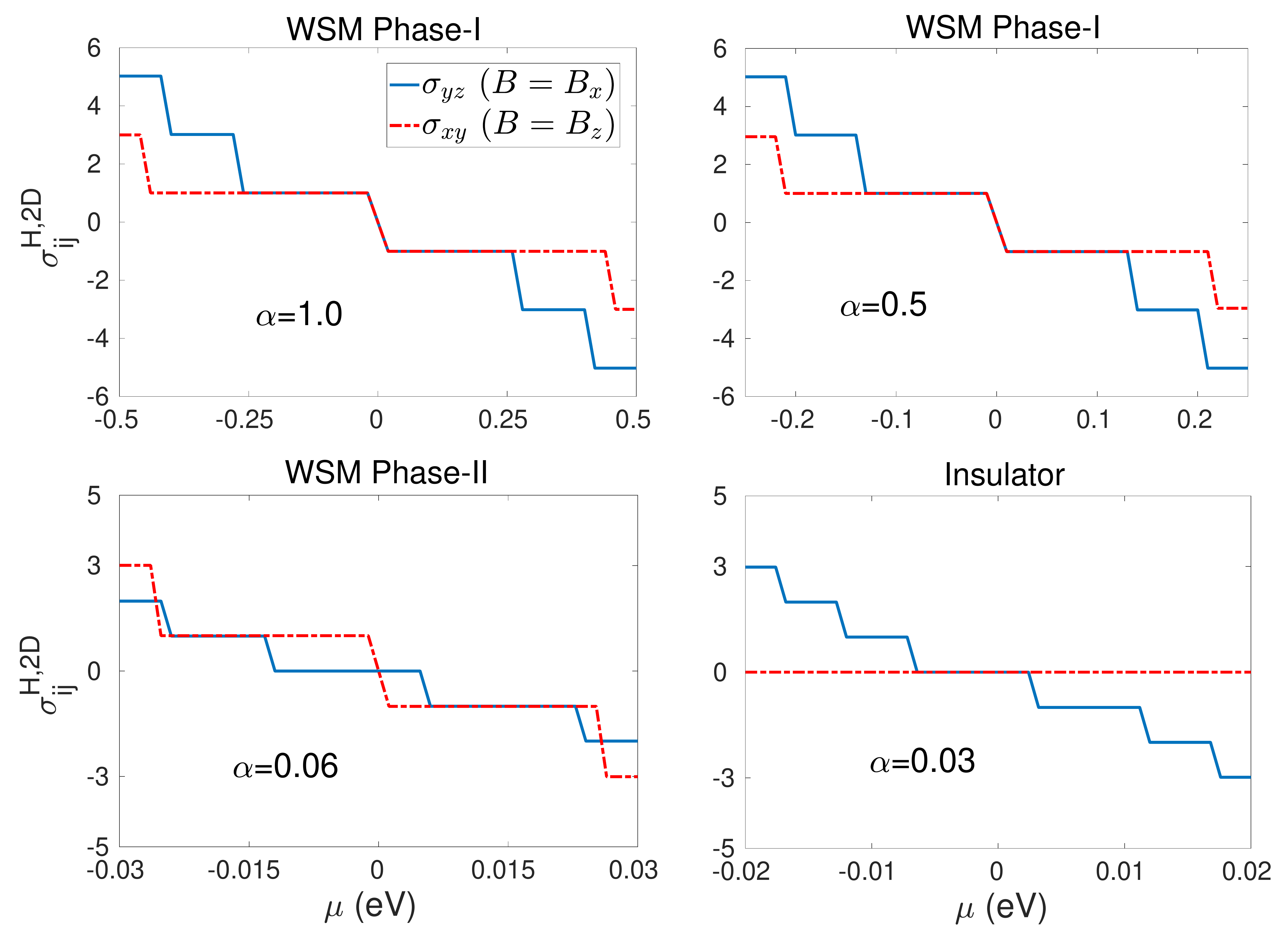}
\caption{The 2D SHC (in unit of $e^2/h$) of WH system as a function of $\mu$ for different strengths of $\rm U$ is depicted. The blue solid line and red dash-dot line represent the SHC when $B$ is applied along $z$ and $x$ directions respectively. We fix $k_x=\cos^{-1}(m/2t_{\rm nor})$ for WSM phase-I and $k_x=\cos^{-1}(m/2t_{\rm nor} -2)$ for WSM phase-II as well as insulating phase to obtain $\sigma_{yz}^{H,2D}$. We keep $k_z=0$ for all the phases to calculate $\sigma_{xy}^{H,2D}$. Here, we have chosen $L_y=70$, $B_x=B_z=2\pi/L_y$ and all the other parameters are same as Fig.~\ref{fig2}.}\label{fig3}
\end{center}
\end{figure}
\section{Quantum Hall Effect} To demonstrate the possible experimental signatures of the topological phase transitions as a function of $\rm U$, we calculate the Hall conductivity using the Kubo linear-response theory, which can be expressed as:
\begin{eqnarray}
\sigma_{ij}^{H}=\frac{ie^2}{hN}  \sum_{\alpha, \beta \neq \alpha} \frac{f_{\alpha}-f_{\beta}}{\epsilon_{\alpha}-\epsilon_{\beta}}\frac{\langle \psi_{\alpha}| v_i |\psi_{\beta} \rangle \langle \psi_{\beta}| v_j |\psi_{\alpha} \rangle}{(\epsilon_{\alpha}-\epsilon_{\beta}+i\delta)}, 
\label{QHE_Kubo}
\end{eqnarray}
where $f_{\alpha}$ denotes the Fermi-Dirac distribution function, $\epsilon_{\alpha}$ represents the eigenvalue of the eigenstate $|\psi_{\alpha} \rangle$, $N=n_x \, n_z$ is the normalization factor ($n_x$ and $n_z$ are the lengths of the system along $x$ and $z$ directions respectively), $v_i=\frac{\partial H}{\partial k_i}$ is the velocity operators, and disorder is included via the level broadening factor $\delta$, i.e., $\delta \rightarrow 0$ indicates clean system. It is
important to note that in the current work, within the chosen
Landau gauge such that the translation invariance along the
$y$ direction is broken, the system Hamiltonian can be represented by a mixed basis of the momentum space (i.e., $k_x$ and
$k_z$) and the real space (i.e., $y$ direction). Then the eigenindex $\alpha$ in the general Hall conductivity formula described by
Eq.~(\ref{QHE_Kubo}) can be decoupled into ($k_x$, $k_z$, and $\alpha$) with $\alpha$ the
site index corresponding to the $y$ direction in real space. We first investigate the two-dimensional sheet Hall conductivity (SHC) $\sigma_{ij}^{2D} (k_l)$  with $i \neq j \neq l$, which can be obtained from the Eq.~(\ref{QHE_Kubo}) by summing over only the momenta parallel to $B$, with dimensionality $e^2/h$. Then the 3D Hall conductivity (QHC) can be written as $\sigma_{ij}^{3D} =\sum_{k_l} \sigma_{ij}^{2D} (k_l)/n_l$ with dimensionality $e^2/h$ per length, where $n_l$ is the length along the $l$-direction. 

Figure~\ref{fig3} presents the SHC ($\sigma_{ij}^{H,2D}$) as a function of doping $\mu$ for various values of $\alpha$. In this work, we restrict $\mu$ to lie within the bulk gap of achiral LLs to clearly examine the contribution of the chiral LLs to the SHC signal. It is clear from Fig.~\ref{fig3} that when $\mathbf{B}$ is applied along the vector connecting Weyl nodes (i.e., $x$-direction), $\sigma_{yz}^{H,2D} (k_x)$ in WSM phase I and II, and the insulating phase, exhibit a quantized staircase profile with quantization changes when $\mu$ crosses from one $k_z$-independent flat LL to another. In the bulk gap of achiral LLs, the SHC is purely composed of chiral LLs in the various WSM phases, whereas the SHC vanishes within the gap for the insulating phase due to the absence of chiral LLs. The width of the plateau of the SHC is determined by the gap size between two consecutive flat LLs. Interestingly, in WSM phase-I the quantization of the SHC changes in steps of $\pm 2$ due to the two-fold degeneracy of LLs, whereas in WSM phase-II the SHC jumps by $\pm 1$ since the LLs are non-degenerate. This fact allows us to track the phase transition between Weyl phases.

In the case of $\rm B$ applied perpendicular to the vector connecting Weyl nodes, i.e., $z$-direction, the SHC $\sigma_{xy}^{H,2D} (k_z)$ in both Weyl phases displays a similar staircase profile, but with quantization steps of $\pm 2$ in both Weyl phases in contrast to the $\mathbf{B}=B_x\hat{x}$. We note that $\sigma_{ij}^{H,2D}$ in WSM phase-I is symmetric about $\mu=0$ ($\sigma_{ij}^{H,2D} (k, \mu)=-\sigma_{ij}^{H,2D} (k, -\mu)$) due to particle-hole symmetric LL spectrum. On the other hand, the above relation does not hold in WSM phase-II, specifically when $\mathbf{B} \parallel \hat{x}$, $\sigma_{yz}^{H,2D} (k_x, \mu) \neq  -\sigma_{yz}^{H,2D} (k_x, -\mu)$ due to the asymmetric nature of the flat
LL spectrum. This striking sensitive dependence on the $B$ direction allows us to distinguish different Weyl phases. The SHC profile we obtained as a function of doping for fixed $\rm{B}$ may also be realized by varying the magnetic field with $\mu$ kept fixed. We would like to point out that the $k$-resolved 3D WH system can be thought as an effective 2D system. This implies the 2D sheet longitudinal conductivity ($\sigma_{ii}^{2D}$) will be non-vanishing analogous to the integer quantum Hall regime in pure 2D systems, specifically, showing peaks with one-to-one correspondence to the step jumps in SHC only when the Fermi energy is within a Landau band where the back-scattering process are present. To obtain the usual peak structure of $\sigma_{ii}^{2D}$ one can simply include a small random onsite disorder in the Hamiltonian to broaden the LLs and inducing a minimal effect on the staircase profile of SHC, but we leave this to a future study. 

\begin{figure}[htb]
\centering
\begin{center}
\includegraphics[width=0.49\textwidth]{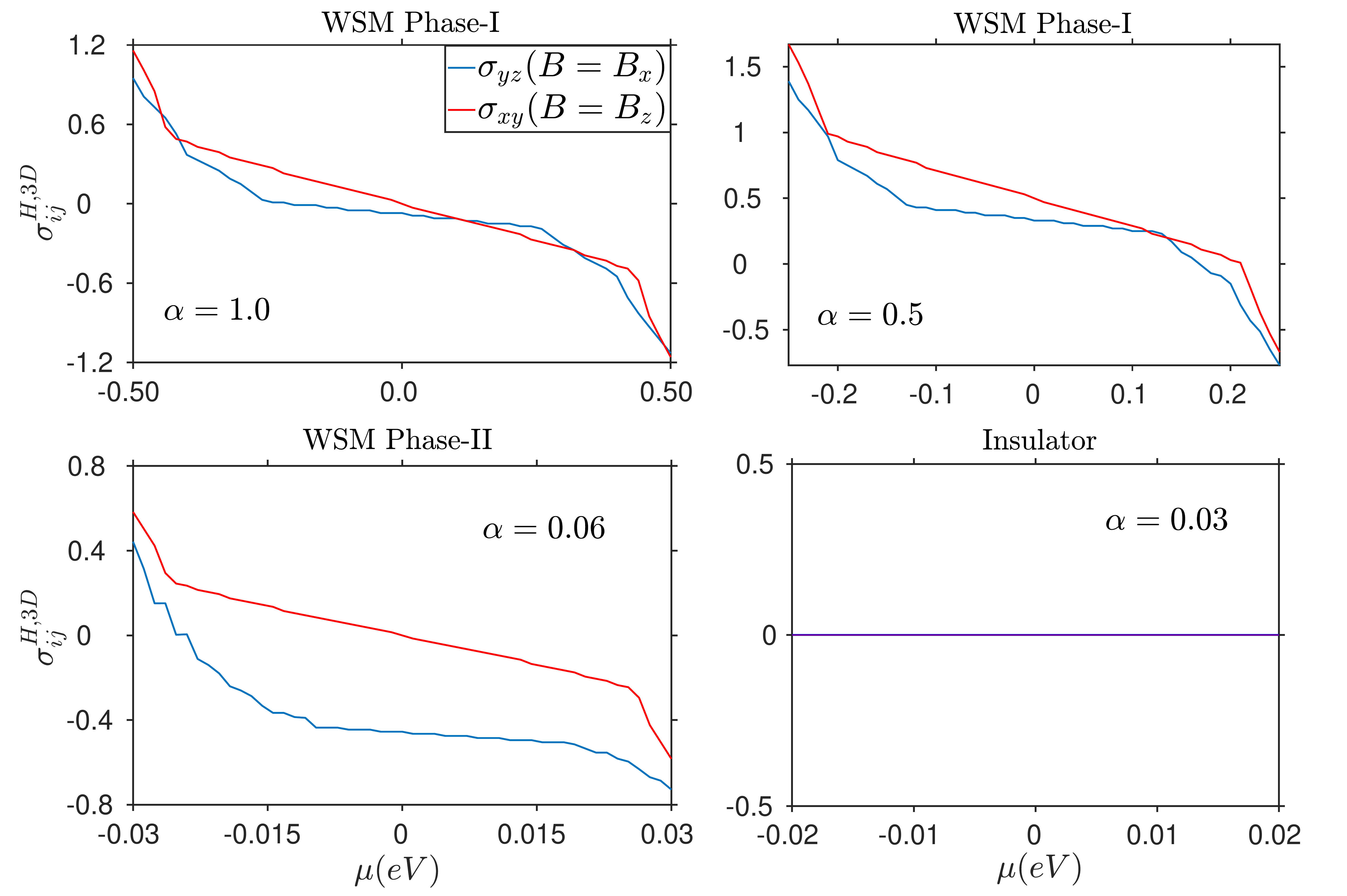}
\caption{The 3D QHC (in unit of $e^2/h  n_l$) of WH system as a function of $\mu$ for different phases is depicted. The blue solid line and red dash-dot line represent the 3D QHC when $B$ is applied along $z$ and $x$ directions respectively. Here, we have chosen $L_y=70$, $B_x=B_z=2\pi/L_y$ and all the other parameters are same as Fig.~\ref{fig2}. $n_l$ is the length of the system along the applied magnetic field direction.}\label{fig4}
\end{center}
\end{figure}

Having explained 2D SHC, we now turn our focus on 3D quantum Hall conductivity $\sigma_{ij}^{H,3D}$, which as a function of doping $\mu$ for various values of $\alpha$ are shown in Fig.~\ref{fig4}. The different Chern insulator planes combine to yield the 3D QHC with quantized SHC along $k_x$ direction. It is clear from the Fig.~\ref{fig4} that the 3D QHC does not exhibit a staircase profile structure as observed for 2D SHC. Since there exists $n_x/n_z$ degenerate LLs associated with each perpendicular momentum mode $k_z/k_x$, after the summation, the quantization is destroyed due to interference among various $\sigma_{ij}^{H,2D}$ ($k_z/k_x$) profiles. We find that the 3D QHC varies linearly with $\mu$ within the bulk gap of achiral LLs indicating solely the chiral LLs contribution in both WSM phases and vanishes in insulating phase. However, when $\mu$ is varied outside the bulk gap of achiral LLs, the 3D QHC follows a nonlinear behavior in $\mu$ due to the admixture of bulk LLs, thereby destroying the linear behavior. The particle-hole asymmetric LL spectrum is inherited from its 2D SHC components for WSM phase-II, see Fig.~\ref{fig4}. Moreover, the slope of 3D QHC  within the bulk gap of achiral LLs increases as we change the magnetic field direction from $x$-axis to $z$-axis in both Weyl phases. We further note that the magnitude of the 3D QHC is decreasing as we tune the system from the weakly interaction regime (WSM phase-I) to the strongly correlated phase (WSM phase-II) due to the flattening of the band dispersions and concomitantly reducing the band velocity.

\section{Discussions} In summary, we study both 2D SHC and 3D QHC in a model Weyl-Hubbard system with both IS and TRS broken to explore the effect of onsite Hubbard correlations on nontrivial Weyl band topology in the presence of an external B-field. Interestingly, along with narrowing bandwidth, we find the chiral LLs change sign from WSM phase-I to WSM phase-II. We calculate the magnetic-field dependent quantum Hall conductivity within the Kubo response theory which shows distinct signatures of topological phase transitions and correlation effects. In particular, the 2D SHC in both Weyl phases depicts staircase profile as a function of doping and displays a qualitatively different quantization between two Weyl phases. However, since the 3D QHC is constructed from the sum of interfering 2D SHCs, it does not show any quantized profile. Interestingly, a linear-$\mu$ behavior appears within the bulk gap of achiral LLs of WSM phases due to chiral LLs where its slope can be quantitatively changed by the changing the magnetic field direction. The recently proposed correlated magnetic WSMs such as $\rm Co_3Sn_2S_2$, $\rm Pr_2Ir_2O_7$ can be the candidate materials to verify the behavior of QHC obtained in this work directly in experiments. We would like to point out that the
strong correlation effect is determined by the ratio of onsite Coulomb interaction to the hopping parameter, $U/t$. Since the
onsite Coulomb interaction $U$ is local, it is relatively difficult
to tune this quantity in experiment. Therefore, to access the
Weyl phase II, a realistic Weyl system with a reduced kinetic
energy (i.e., hopping) is desirable, which can be achieved via the effective negative pressure.

For a typical WSM, the lattice constant $a \sim 1 \, \rm{nm}$. Therefore, considering the length of the sample $L=100\,a$ gives the magnetic field strength $B \sim 41 \, \rm{T}$. In this case, the magnetic length turns out to be $l_c \sim 4 \, \rm{nm}$ which satisfies the condition $L>>l_c>>a$ (away from the Butterfly regime) and also within the well reach in experimental feasibility. We would like to point out that the surface Fermi arc contribution can also be important to QHC. However, in the present study, this contribution is negligible due to following reason: it has been shown that when the external $\rm{B}$ is greater than $\rm{B_{sat}}$ where $\rm{B_{sat}}=k_0/L$ with $k_0$ is the arc length of the Fermi arc, the majority of the magnetic cyclotron orbit takes place in the bulk and the surface Fermi arc contribution becomes negligibly small~\cite{Potter_2014}. In the present case, considering a thick slab of WSM with $L>>l_{c}$ and $B(=2\pi m /L)>B_{sat}$ leading to the fact that the QHC will be dominated by bulk chiral and achiral LLs. In addition, periodic-boundary condition along the y-direction might also reduce finite size effects~\cite{Nag_2022}. However, investigating the Fermi arc contribution to QHC in detail is an important question we leave for future studies.

Overall, our study demonstrates that the QHE persists even in the presence of strong electron-electron interactions and provides distinct signatures of different topological phases. This make QHE an efficient direct probe of band topology in correlated quantum materials.

\textcolor{blue}{\it Acknowledgements:} The work at Los Alamos National Laboratory was carried out under the auspices of the U.S. Department of Energy (DOE) National Nuclear Security Administration under Contract No. 89233218CNA000001. It was supported by the LANL LDRD Program, and in part by the Center for Integrated Nanotechnologies, a DOE BES user facility, in partnership with the LANL Institutional Computing Program for computational resources.

\bibliography{WH_QHE}{}
\bibliographystyle{apsrev4-2}

\end{document}